\documentclass[aps,prb,showpacs,groupedaddress,superscriptaddress,twocolumn]{revtex4}

\usepackage{graphicx,amsmath}
\usepackage{color}
\begin{document}

\title{Two-Dimensional Lattice Model for the Surface States of Topological Insulators}

\author{Yan-Feng Zhou}
\affiliation{International Center for Quantum Materials, School of Physics, Peking University, Beijing 100871, China}
\affiliation{Collaborative Innovation Center of Quantum Matter, Beijing 100871, China}

\author{Hua Jiang}
\email[]{jianghuaphy@suda.edu.cn}
\affiliation{College of Physics, Optoelectronics and Energy, Soochow University, Suzhou 215006, China}

\author{X. C. Xie}
\affiliation{International Center for Quantum Materials, School of Physics, Peking University, Beijing 100871, China}
\affiliation{Collaborative Innovation Center of Quantum Matter, Beijing 100871, China}

\author{Qing-Feng Sun}
\email[]{sunqf@pku.edu.cn}
\affiliation{International Center for Quantum Materials, School of Physics, Peking University, Beijing 100871, China}
\affiliation{Collaborative Innovation Center of Quantum Matter, Beijing 100871, China}

\date{\today}

\begin{abstract}
The surface states in three-dimensional (3D) topological insulators can be described by a two-dimensional (2D) continuous Dirac Hamiltonian.
However, there exists the Fermion doubling problem when putting the continuous 2D Dirac equation
into a lattice model.
In this paper, we introduce a Wilson term with a zero bare mass into the 2D lattice model to overcome the difficulty.
By comparing with a 3D Hamiltonian, we show that the modified 2D lattice model can faithfully describe the low-energy electrical and transport properties of surface states of 3D topological insulators.
So this 2D lattice model provides a simple and cheap way to numerically simulate
the surface states of 3D topological-insulator nanostructures.
Based on the 2D lattice model, we also establish the wormhole effect in a topological-insulator nanowire
by a magnetic field along the wire and show the surface states being robust against disorder.
The proposed 2D lattice model can be extensively applied to study the various properties
and effects, such as the transport properties, Hall effect, universal conductance fluctuations,
localization effect, etc..
So it paves a new way to study the surface states of the 3D topological insulators.
\end{abstract}


\maketitle

\section{\label{sec1}Introduction}

The three-dimensional (3D) topological insulators (TIs), which possess both insulating bulk states and conducting surface states protected by time-reversal symmetry, have attracted intense attention\cite{HMZ,Qxl}.
The surface states present an odd number of gapless Dirac cones and are robust against time-reversal-invariant disorders due to their spin-momentum locked helical properties.
The surface states have been confirmed by the angle-resolved photoemission spectroscopy (ARPES) experiments\cite{HD,Chen,Xia}.
However, it is a challenge to observe quantum transport of the surface states , which are usually covered
by bulk carriers caused by material defects\cite{ButchN}.
Recently, several unambiguous experimental observations of quantum Hall effect based on surface states have been obtained in TI materials\cite{YXu1,RYoshimi,NKoirala,YXu2}.
Another experimental evidence for the surface transport is the Aharonov-Bohm conductance oscillations in a TI nanowire with a magnetic field paralleling the wire\cite{HPeng,FXiu,Dufouleur,SHong,SCho}.
The oscillations are subject to an anomalous phase shift of half flux quanta $\phi_0/2=h/2e$
arising from a $\pi$ Berry phase for surface states on a curved surface in 3D TI nanowires\cite{GRosenberg,JBardarson,REgger,YZhang}.

On the other hand, to quantitatively simulate the transport properties of surface states, a low-energy lattice model is desirable to describe the thin films or nanostructures of 3D TIs.
The 3D TIs can be described by a continuous 3D effective model with parameters that obtained by fitting the energy spectrum from \emph{ab initio} calculations\cite{HZhang,CXLiu,WYShan}.
By discretizing the 3D effective model, a 3D cubic lattice Hamiltonian can be a candidate for further calculations.
Moreover, a 3D Fu-Kane-Mele model in diamond lattice with spin-orbit interactions can also realize 3D TI phase by choosing suitable parameters\cite{LFu}.
However, it is computationally expensive to do calculations basing on the Hamiltonian derived from a 3D lattice model. To acquire a high-accuracy result, a
huge-dimension Hilbert space is needed. Despite of insulating bulk states, the low-energy description of surface states in 3D TIs
around the Dirac node is solely given by a massless 2D Dirac Hamiltonian.
The 2D lattice model can greatly reduce the computational cost with a same precision in comparison with 3D lattice model. However, there exists the Fermion doubling problem, when placing the massless 2D Dirac equation into a lattice form\cite{Nielsen,Kogut,Vafek2}.
As shown in Fig.1(c), three additional Dirac nodes appear at boundaries of the first Brillouin zone.
This has impeded the utility of a 2D lattice model to quantitatively study the fascinating transport properties related to surface states in 3D TIs.

In this paper, we propose a square-lattice model in terms of 2D Dirac Hamiltonian with a Wilson term
to describe the surface states of 3D TIs.
We show that the modified 2D lattice model can mimic the linear dispersion of surface states near the $\Gamma$ point and opens energy gaps at the spurious Dirac cones located at boundaries of the Brillouin zone (see Fig.1).
In particular, its Berry phase is very close to $\pi$ for the low Fermi level case, which is
well consistent with 3D model.
The proposed 2D model can be extensively applied to study the various properties
and effects of the surface states of the 3D topological insulator, such as the transport properties,
Hall effect, universal conductance fluctuations, localization effect, band structures, and so on.
Moreover, as an example of the application, the electrical and transport properties of a cuboid TI nanowire
are studied by employing the proposed 2D lattice model.
The results show that the surface spectrum opens a small gap and is doubly degenerate, which is same with the band structure calculated from 3D model.
From the 2D lattice model, we also establish the wormhole effect by threading a magnetic
flux $\phi$ across the cross section of TI nanowire.
Further, we obtain a consistent picture for the conductance of a clean TI nanowire from both the 2D and 3D lattice models and show that the conductance is robust against the surface disorders.
All these investigations show that the proposed 2D lattice model
can faithfully describe the surface states of 3D TI in the long wavelength limit.
Note that the numerical calculation based on the 2D lattice model is much quicker than the one based
on 3D model.

The rest of the paper is organized as follows. In Sec.~\ref{sec2}, the theoretical model and the methods for describing the surface states of 3D TI are presented. In Sec.~\ref{sec3}, we demonstrate the utility of the proposed 2D lattice model by studying the band structure and transport properties of a cuboid TI nanowire.
Sec.~\ref{sec4} discusses the advantages of the proposed 2D lattice model.
Finally, a brief summary is given in Sec.~\ref{sec5}.

\section{\label{sec2}Model and Methods}

\begin{figure}
\includegraphics[scale=0.3]{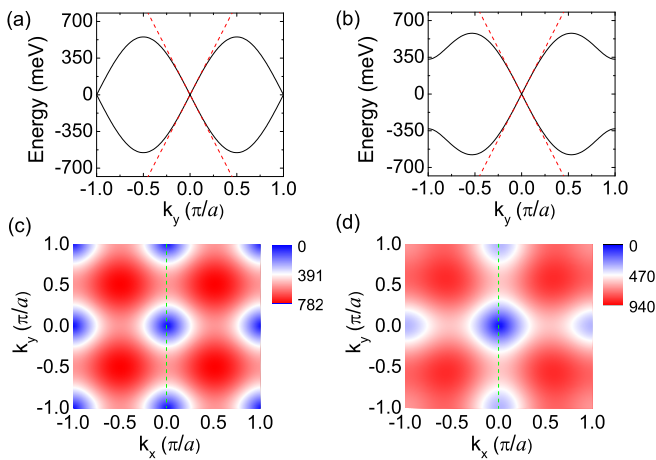}
\caption{ (Color online) Band structure for (a) Hamiltonian $H_0$ in Eq.(\ref{eq:2})
and for (b) Hamiltonian $H=H_0+H_W$ with $W=0.3\hbar\nu_F$ along the path indicated by green dash line in the first Brillouin zone in (c,d). The red dash line in (a,b) corresponds to band structure of continuous Hamiltonian in Eq.(\ref{1}).
(c) and (d) show 2D maps of the conduction bands of Hamiltonians $H_0$ and $H$ in the whole first Brillouin zone, respectively.}
\end{figure}

The low-energy 2D effective Hamiltonian for the surface states of 3D TI is\cite{LHZ,LiuH},
\begin{equation}\label{1}
  H({\bf k}) = \hbar \nu_{F}(\hat{\sigma}\times {\bf k})\cdot \hat{n},
\end{equation}
where $\nu_F$ is the Fermi velocity, $\hat{\sigma} \equiv (\sigma_x,\sigma_y,\sigma_z)$ with $\sigma_{x,y,z}$ the Pauli matrices, ${\bf k}=(k_x,k_y,k_z)$ is the momentum, and $\hat{n}$ is the normal vector of a specific surface. Its spectrum has a single Dirac cone construction with $\epsilon_{\bf k}=\pm\hbar\nu_{F}\vert {\bf k}_n\vert$, in which ${\bf k}_n$ is the component of $\bf k$ perpendicular to $\hat{n}$. For simplicity, we take $\hat{n}$ parallel to z direction and ${\bf k}_n=(k_x,k_y)$ in the following. Now, we discretize the Hamiltonian in Eq.(\ref{1}) to obtain a square lattice Hamiltonian\cite {book1},
\begin{equation}\label{eq:2}
\begin{split}
&H_0 = \sum \limits_{\mathrm{\bf{i}}}\frac{i\hbar \nu_{F}}{2a}(c_{\mathrm{\bf{i}}}^{\dag}\sigma_{y}c_{\mathrm{\bf{i}}+\delta\hat{\bf{x}}} -c_{\mathrm{\bf{i}}}^{\dag}\sigma_{x}c_{\mathrm{\bf{i}}+\delta\hat{\bf{y}}})+H.c.,\\
\end{split}
\end{equation}
where $c_{\mathrm{\bf{i}}}$ and $c_{\mathrm{\bf{i}}}^{\dag}$ are the annihilation and creation operators at site $\mathrm{\bf{i}}$, respectively. $\delta\hat{\bf{x}}$ ($\delta\hat{\bf{y}}$) is the primitive vectors of quare lattice along the x (y) direction, and $a$ is the lattice constant.
In this paper, we set the Fermi velocity $\nu_{F} = 5\times10^{5}\ \rm m/s$ and the lattice constant $a = 0.6\ \mathrm{nm}$.
This Fermi velocity $\nu_F$ corresponds to topological insulator $\mathrm{Bi}_2\mathrm{Te}_3$ and $(\mathrm{Bi}_{x}\mathrm{Sb}_{1-x})_2\mathrm{Te}_3$, which can be experimentally extracted from ARPES and scanning tunneling spectroscopy.\cite{RYoshimi,ZhangT}

It can be seen from the band structure shown in Fig.1(a) that the lattice Hamiltonian $H_0$ can reproduce the linear spectrum of the effective Hamiltonian $H({\bf k})$ in Eq.(\ref{1}) near the $\Gamma$ point.
Unfortunately, the lattice model also has low-energy excitations with linear dispersion in the vicinity of points (0,$\pm\pi/a$). We can identify four Dirac cones in the first Brillouin zone [see Fig. 1(c)].
As mentioned above, this is the Fermi doubling problem in a lattice model for chiral Weyl fermions and involved with continuous chiral symmetry\cite{Nielsen}. In order to avoid the Fermi doubling problem, we introduce a Wilson term with a zero bare mass into the lattice Hamiltonian\cite{Kogut,Vafek2}:
\begin{equation}\label{eq:3}
\begin{split}
&H_{W} = \sum \limits_{\mathrm{\bf{i}}}(2W/a)c_{\mathrm{\bf{i}}}^{\dag}\sigma_{z}c_{\mathrm{\bf{i}}}\\
&-\sum\limits_{\mathrm{\bf{i}}}(W/2a)[(c_{\mathrm{\bf{i}}}^{\dag}\sigma_{z}c_{\mathrm{\bf{i}}
+\delta\hat{\bf{x}}}+c_{\mathrm{\bf{i}}}^{\dag}\sigma_{z}c_{\mathrm{\bf{i}}+\delta\hat{\bf{y}}})]+H.c.,\\
\end{split}
\end{equation}
which is the counterpart of a continuous term $(Wa/2)k_n^2\sigma_z$.
In fact, the Wilson term which includes the Pauli matrix $\sigma_z$
acts as a momentum-dependent mass term and breaks the chiral symmetry explicitly.
For momentum near the $\Gamma$ point, i.e., the center of Brillouin zone, the Wilson term vanishes quadratically, and thus the spectrum keeps gapless, in the long wavelength limit.
However, for the doubler fermions, that are the states at boundaries of the first Brillouin zone, the Wilson term is nonvanishing and opens a finite gap.
One also can obtain the same conclusion from the spectrum of the lattice Hamiltonian.
The spectrum of Hamiltonian $H=H_0+H_W$ in a lattice can be formulated by  $E_{k_x,k_y}^2=\frac{\hbar^2\upsilon_F^2}{a^2}\sum_{i=x,y}\sin^2(k_ia)+\frac{4W^2}{a^2}(\sum_{i=x,y}\sin^2(k_ia/2))^2$. For $k_ia\rightarrow (0,0)$, $E^2$ approaches to $\hbar^2\upsilon_F^2k_n^2+W^2a^2k_n^4/4$, which gives a linear dispersion at the low energy case.
The Wilson term can open an energy gap $\Delta_1=4W/a$ at points (0,$\pm\pi/a$) and ($\pm\pi/a$,0) [Fig.1(b)] and a gap $\Delta_2=8W/a$ at points $(\pm\pi/a,\pm\pi/a)$,
so that three additional Dirac cones arising from the Fermi doubling are removed, leaving the single one near the $\Gamma$ point [Fig.1(d)].
On the other hand, if the Fermi level is low (small Fermi momentum $k_F$), the Berry phase for eigenstates $\psi_{\pm}({\bf k})$ of Hamiltonian $H$ around the Fermi surface is $\gamma_{\pm}=\pi(1\pm \frac{Wa}{2\hbar\nu_F}k_F)$ modulo $2\pi$, which is very closed to the $\pi$ Berry phase of surface states in 3D TIs, although the Wilson term has broken the time-reversal symmetry. Therefore, the lattice Hamiltonian $H$ can restore properties of surface states in the long wavelength limit.

\begin{figure}
\includegraphics[scale=0.34]{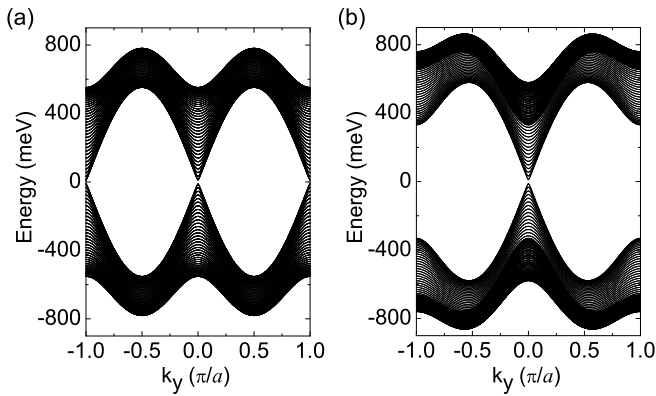}
\caption{Band structure of a cuboid TI nanowire calculated from the effective 2D Hamiltonian $H^{2D}$ in
Eq.(\ref{eq:4}). (a) For $W=0$, each band is fourfold degenerate. (b) For $W=0.3\hbar\nu_F$, all bands are doubly degenerate. The nanowire has a cross section of size $(L_x, L_z)=(23.4 \ \mathrm{nm}, 24.6\ \mathrm{nm}).$ } \label{fig:5}
\end{figure}

\section{\label{sec3}Band structure and transport properties}

Next, we quantitatively study the properties of a cuboid TI nanowire [see Fig.3(a)]
based on the lattice Hamiltonian $H$ proposed above.
The nanowire has a cross section of a size $(L_x, L_z)$ and is invariant under translation along the y axis, so that the momentum $k_y$ is a good quantum number.
To study the surface states of 3D TI nanostructure with non-planar surfaces,
one has to solve the Dirac Hamiltonian in curved 2D spaces\cite{LeeD1,VafekO}.
For a cuboid TI nanowire, one can completely describe the surface states by different Hamiltonians on specified surfaces as provided in Eq. (\ref{1}) supplemented with matching conditions at boundary lines\cite{BreyL,Deb}.
Furthermore, by some local unitary transformations, an effective 2D model can be derived for the four surfaces\cite{TM,TM2,ZhouY}.
For the -x/-z/+x surfaces, we make rotations by fixing the y axis and the original x axis is changed into z/-x/-z axes.
After corresponding unitary transformations, the effective 2D Hamiltonian for surface states reads
\begin{eqnarray}\label{eq:4}
&&H^{2D}=\sum^{N-1}_{n}\sum_{m}[c_{nm}^{\dag}T_{0}c_{nm}+(c_{nm}^{\dag}T_{x}c_{n+1,m} \nonumber\\
&&+c_{nm}^{\dag}T_{y}c_{n,m+1}+H.c.)]-\sum_{m}c_{Nm}^{\dag}T_{x}c_{1m}+H.c.,
\end{eqnarray}
with
\begin{eqnarray}
T_0&=&(2W/a)\sigma_z, \nonumber\\
T_x&=&-(W/2a)\sigma_z+(i\hbar \nu_{F}/2a)\sigma_{y},\nonumber\\
T_y&=&-(W/2a)\sigma_z-(i\hbar \nu_{F}/2a)\sigma_{x},\nonumber
\end{eqnarray}
where $c_{nm}$ and $c_{nm}^{\dag}$ are the annihilation and creation operators at site $(n,m)$ respectively. N is the total number of lattices encircle the TI nonowire.

To demonstrate the effectiveness of this 2D lattice Hamiltonian, we firstly calculate the band structure of surface states for a 3D TI nanowire [see Fig.2].
Without the Wilson term ($W=0$), each band is fourfold degenerate and the low-energy spectrum emerges
at both $k_y=0$ and $\pm\pi/a$ as shown in Fig.2(a).
Two of the quadruplet can be ascribed to the redundant species doubling at boundary with $k_x=\pm\pi/a$ in the first Brillouin zone [see Fig.1(c)].
These band structures, even if at the low energy, are completely different with 3D TI models in Eq.(1) or Eq.(\ref{eq:5}).
However, by introducing the Wilson term with $W=0.3\hbar\nu_F$,
it can be clearly observed in Fig.2(b) that the redundant degenerate modes are shifted away from the doubly degenerate ones. Now each band is twofold degenerate due to the same eigenvalues for surface states on the opposite surface. A large gap is opened at $k_y=\pm\pi/a$ and the low-energy spectrum only involves in the vicinity of $\Gamma$ point. Here a small gap is also opened at $k_y=0$ because of the $\pi$ Berry phase for carriers taking a circle around the nanowire.

Here, it is worth noting that the coefficient $W$ in the Wilson term needs to be fine tuned in the calculation.
For the smaller $W$, the band structure of the Hamiltonian $H^{2D}$ can perfectly be coincident with that of the 3D model $H^{3D}$, but the gap $\Delta_1=4W/a$ due to the Wilson term at the boundary of first Brillouin zone may be too small to settle the Fermion doubling problem. On the other hand, the bigger $W$ can well settle the Fermion doubling problem, but it may caused a serious departure near the $\Gamma$ point.
So $W$ can not be too big and too small.
While $W$ in the range of $0.3\hbar\nu_F$ to $1.0\hbar\nu_F$,
the band structures of the Hamiltonian $H^{2D}$ can perfectly be coincident with that of
$H^{3D}$ from about $-200 \ \mathrm{meV}$ to $200 \ \mathrm{meV}$, and the gap is large enough at the boundary of first Brillouin zone also.
This means that our method can work well in a large range of $W$.

To further ensure the accuracy of the 2D lattice Hamiltonian in Eq.(\ref{eq:4}),
we make a comparison with the low-energy approach of Zhang et al. \cite{HZhang}, in which the Hamiltonian near the $\Gamma$ point has the form:
\begin{equation}\label{eq:5}
\begin{split}
&H^{3D}=\epsilon_0({\bf k})\sigma_0\tau_0+M({\bf k})\sigma_0\tau_z+A_1k_z\sigma_z\tau_x\\
&+A_2(k_x\sigma_x+k_y\sigma_y)\tau_x,\\
\end{split}
\end{equation}
with $k_{\perp}^2=k_x^2+k_y^2$, $\epsilon_0({\bf k})=C+D_1k_z^2+D_2k_{\perp}^2$, and $M({\bf k})=M_0-B_1k_z^2-B_2k_{\perp}^2$. $\tau_{x,y,z}$ are Pauli matrices in orbital space and $\sigma_0$/$\tau_0$ denote the identity matrix. In this paper, we have made further simplification that $\epsilon_0({\bf k})=0$, $M_0=0.28\ eV$, $B_1=B_2=0.1\ \mathrm{eV \cdot nm^2}$, and $A_1=A_2=\hbar\nu_F=0.33\ \mathrm{eV \cdot nm}$.
Here the parameters $M_0$ and $B_1$ are assigned to the same values
with $\mathrm{Bi}_2\mathrm{Se}_3$,\cite{HZhang} and the Fermi velocity $\nu_F$ is the same with Fig. 1.
These parameters guarantee the Hamiltonian with isotropy and particle-hole symmetry.
By discretizing the Hamiltonian (\ref{eq:5}),
we obtain a cubic lattice model by which surface states can be derived with open boundary conditions\cite{HJiang}.
Fig.3(b) plots the band structure calculated from both the 2D and 3D lattice models.
The two different lattice models provide very unanimous results for the spectrum in bulk gap at low energy. So the 2D lattice model proposed above is very effective
for quantitative studies of the surface properties of 3D TI.

\begin{figure}
\includegraphics[scale=0.3]{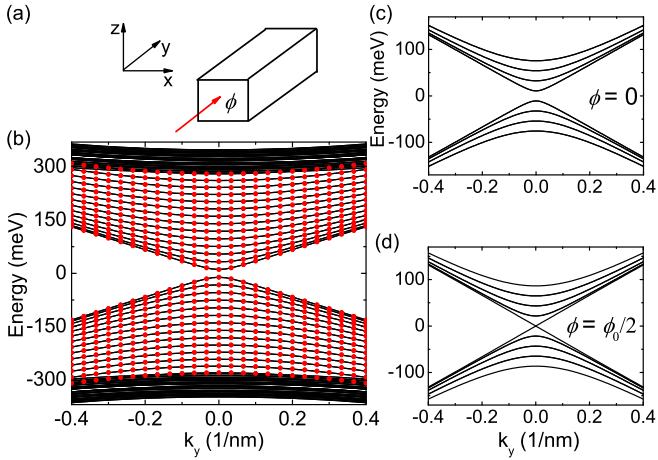}
\caption{ (Color online) (a) Schematic of a cuboid TI nanowire with a magnetic flux $\phi$ across its section cross. (b) Band structure of a nanowire without
magnetic flux, in which the black solid lines are obtained from a 3D lattice Hamiltonian in Eq.(\ref{eq:5}) and the red solid circles are obtained from a 2D lattice Hamiltonian in Eq.(\ref{eq:4}).
(c,d) show the wormhole effect. For $\phi=0$ the spectrum is gapped and doubly degenerate, while for $\phi=\phi_0/2$ a pair of linear modes are present. The parameters are same with Fig.2(b).}
\end{figure}

Now, we show the wormhole effect based on the 2D lattice model of Eq.(\ref{eq:4}).
Since the spin-momentum locked properties of surface states, surface electron obtains a $\pi$ Berry phase while it goes around the four facets of a TI nonowire\cite{GRosenberg,JBardarson,REgger,YZhang}.
The extra $\pi$ Berry phase yields a gapped spectrum of surface state in a TI nanowire as shown in Fig.3(c).
By threading a magnetic flux $\phi$ along the nanowire [see Fig.3(a)],
surface electron also get an Aharonov-Bohm phase\cite{AronovA}.
Here, the effect of longitudinal magnetic field is included by adding a phase term $\phi_{n,n+1}=\int_{n}^{n+1}{\bf A}\cdot d{\bf l}/\phi_0$ to $T_x$ in Eq.(\ref{eq:4}), where ${\bf A}=(0,0,B_yx)$ is the vector potential for a magnetic field $B_y$ parallel to the y direction.
By a flux $\phi=\phi_0/2$, the $\pi$ Aharonov-Bohm phase exactly cancels the $\pi$ Berry phase and a pair of non-degenerate linear modes emerge with closing the gap [Fig.3(d)]. The phenomenon is known as wormhole effect\cite{GRosenberg}.

\begin{figure}
\includegraphics[scale=0.46]{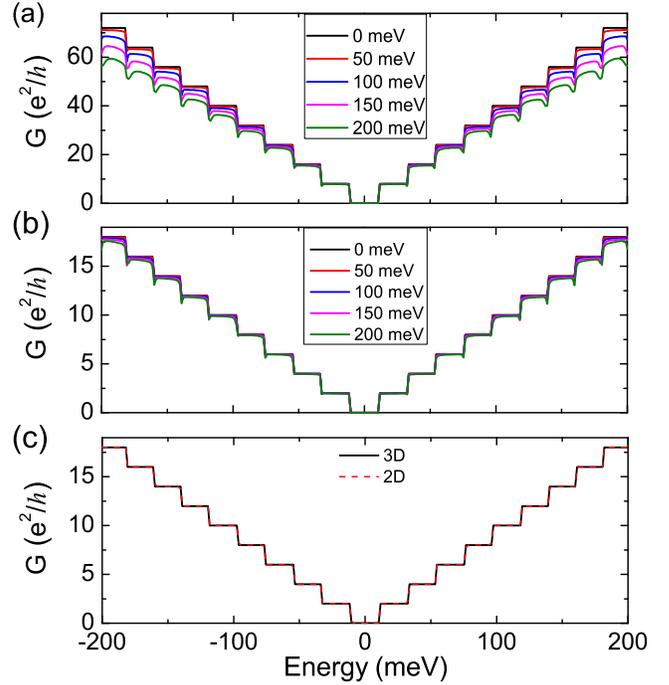}
\caption{ (Color online) Conductance $G$ of the TI nanowire
as a function of Fermi energy $E_F$ calculated by the 2D lattice model in Eq.(\ref{eq:4}) at the different disorder strength $U$ for (a) without and (b) with the Wilson term [$W=0$ in (a) and $W=0.3\hbar\nu_F$ in (b)].
(c) Conductance calculated by both a 3D lattice model (black solid line) and a 2D lattice model (red dash line) with $W=0.3\hbar\nu_F$ for a clean nanowire.
The length of the center region $L_y=24 \ \mathrm{nm}$ and a cross section $(L_x, L_z)=(23.4 \ \mathrm{nm}, 24.6\ \mathrm{nm})$. The curves in (a) and (b) are averaged over up to 100 random configurations.} \label{fig:4}
\end{figure}

Next, we study the effectiveness of the 2D lattice Hamiltonian on the simulation of transport properties.
Here, we construct a two-terminal device by dividing an infinite TI nanowire into a center region with a length $L_y$ and two semi-infinite left/right leads.
The conductance can be calculated from the Landauer-B\"{u}ttiker formula, $G(E_F)=(e^2/h)T_{LR}(E_F)$.\cite{book1}
Here the transmission coefficient $T_{LR}(E_F)=Tr[{\bf \Gamma}_L{\bf G}^r{\bf \Gamma}_R{\bf G}^a]$, in which the ${\bf \Gamma}_{L/R}=i[{\bf \Sigma}_{L/R}^r-{\bf \Sigma}_{L/R}^a]$, the Green function $G^r(E_F)=[G^a]^{\dag}=[E_F-{\bf H}^{\mathrm{cen}}-\sum_{p=L,R}{\bf \Sigma}_p^r]^{-1}$,
with $E_F$ the Fermi energy, and ${\bf H}^{\mathrm{cen}}$ being the Hamiltonian of center region\cite{sun1}.
The retarded self-energy ${\bf \Sigma}_{L/R}^r$ stems from coupling to the left/right lead\cite{LeeD2}.
With the presence of disorders, on each site the term $T_0$ in Eq.(\ref{eq:4}) is changed to $T_0+w_i\sigma_0$, where $w_i$ is uniformly distributed in the range $[-U/2,U/2]$ with disorder strength $U$.
Here, we only consider surface disorders in the center region.
Fig.4(a) and (b) show the conductance as a function of Fermi energy $E_F$ at different disorder strength $U$ without and with the Wilson term, respectively.
For $U=0$, the conductance exhibits quantum plateaus and
the results obtained from a 2D lattice model with a Wilson term agree well with the ones from a 3D lattice model while the Fermi energy is inside the bulk gap [see Fig.4(b) and (c)].
However, without the Wilson term, the 2D lattice model gives a quadrupled conductance owing to additional transport modes from Fermi doubling [Fig.4(a)].
Moreover, for $U\neq0$, the disorders can induce scattering between the quadrupled Dirac cones and the plateau structure is distinctly destroyed for disorder strength $U> 50\ \mathrm{meV}$ [see Fig.4(a)].
As the Wilson term has eliminated the redundant Dirac cones, the quantum plateaus of conductance are robust against disorder [see Fig.4(b)], which is reasonable for surface state in 3D TIs.
Considering that the 2D lattice model proposed in this paper has discarded bulk states, it can be used to simulate the low-energy transport properties of surface states in 3D TI systems with a large size or high accuracy in a much less expensive way.

\section{\label{sec4}Advantages of the 2D lattice model}

In this section, we will discuss the advantages of the proposed 2D lattice model comparing
with other methods and some potential applications. The 2D lattice model is a simplification of 3D model.
In principle, it can not predict phenomena beyond the 3D model.
However, in the practical calculations, the 2D lattice model can greatly improve computation
speed and reduce the memory usage.
This means that our method can deal systems with large size beyond the ability of 3D model
and new physics may appear with increasing the size of samples.
So it is more efficient than other methods in the literatures for numerical simulation.
What's more, the proposed 2D lattice model can be extensively utilized to study
the various properties and effects of the surface states of the 3D topological insulator.
For example, by using this 2D lattice model, one can study the transport properties, Hall effect,
universal conductance fluctuations, localization effect, band structures, and so on.

\begin{figure}
\includegraphics[scale=0.3]{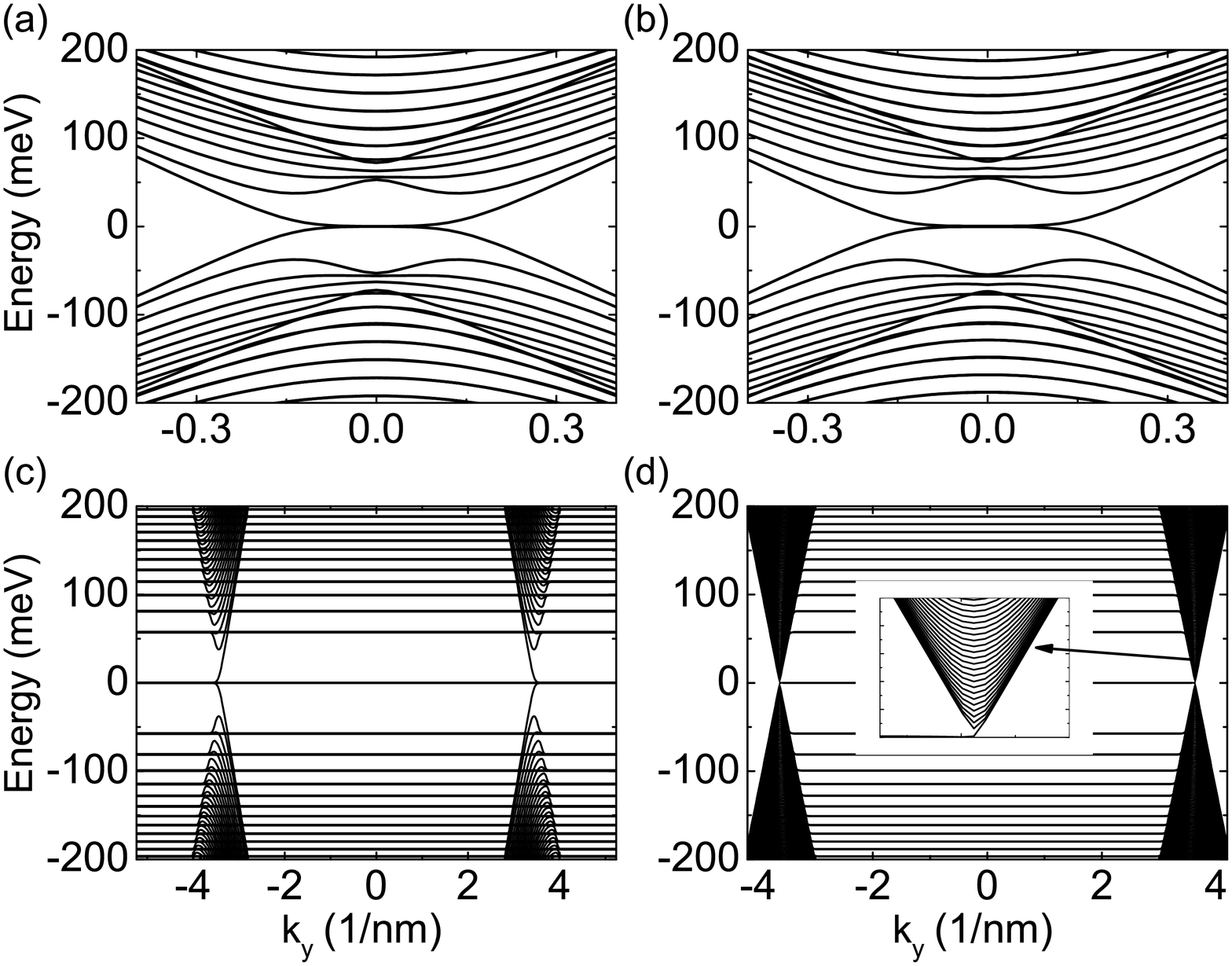}
\caption{
Band structures calculated by (a) 3D lattice model in Eq.(5) and (b) 2D lattice model in Eq.(4)
for a cuboid TI nanowire [see Fig.3(a)] with a cross section $(L_x,L_z)=(23.4 \ \mathrm{nm},24.6 \ \mathrm{nm})$
(the number of lattice being $40\times42$) under a magnetic field along the $z$ direction.
(c) and (d) show band structures calculated by 2D lattice model for
another two large nanowires with a cross sections $1560\times 40$ $(935.4\ \mathrm{nm},23.4\ \mathrm{nm})$
and $800\times 800$ $(479.4 \ \mathrm{nm},479.4 \ \mathrm{nm})$.
The magnetic field $B = 10 \ \mathrm{T}$ and all the other unmentioned parameters
are same with parameters in Fig.2(b).}
\end{figure}

Next, we calculate the band structure in 3D TI samples under a perpendicular
magnetic field to illuminate the fast computing speed and less memory usage.
For a cuboid TI nanowire [see Fig.3(a)] with a cross section $40\times42$
[here $40\times42$ being the number of lattice along the $x$ and $z$ direction
and the corresponding cross section of size $(L_x,L_z)=((40-1)a,(42-1)a)=(23.4 \ \mathrm{nm},24.6 \ \mathrm{nm})$],
the band structure calculated from both 3D lattice model in Eq.(5) and 2D lattice model in Eq.(4)
are shown in Fig.5(a) and (b).
The two models provide very coincident picture for band structure and the zero Landau
level appears as expected.
It is worth noting that only $(40+42+40+42-4)\times2=320$ basis orbits are needed to obtain such results
with the 2D lattice model.
However, it requires $40\times 42\times 4=6720$ basis orbits to work with the 3D lattice model.
So, the new method is more efficient and much quicker.
The ratio of the computation speeds based on the
2D model and 3D model is about $(6720/320)^3 = 9261$.

Moreover, we consider another two large nanowires with a cross
section $1560\times40$ and $800\times800$ [see Fig.5(c) and (d)],
both of which are described with 6392 basis orbits in the 2D lattice model.
The Fig.5(c) clearly shows many unambiguous Landau levels with high index
due to the large size along the $x$ direction.
Moreover, as shown in Fig.5(d), a great many of nonchiral edge modes coexist with chiral edge
mode in the side surface due to the large size in the $z$ direction,
which can destroy the integer quantized Hall plateaus.
These physics are hidden in small size samples because of the quantum confinement.
Notice that it is extremely difficult for 3D lattice model to deal with such large samples.
If by using the 3D lattice model, the basis orbits are
$1560\times40 \times 4 = 249600$ and $800\times800 \times 4=2560000$ for the samples in Fig.5(c) and (d),
and the computation time required are increased by about $5.9\times10^4$ and $6.4\times10^7$ times, respectively.
So, it is impossible to obtain the results of Fig.5(c) and (d) from the 3D model.
Therefore, in practical numerical simulation, the proposed 2D model can give new physics beyond 3D model.

It is also worth mention that although the conductances in Fig.4(c)
from the 3D model in Eq.(5) and 2D model in Eq.(4)
are very consistent, the time used in these two calculations
is very different, the difference is about ten thousand times.
While in the presence of the disorder, we can study the effect of the disorder on the conductances
by using the 2D lattice model [see Fig.4(b)],
but it is very difficult to study this effect based on the 3D model.

Based on the growth of high-quality sample\cite{YXu1,RYoshimi,NKoirala,YXu2},
it becomes a central topic in the study of topological states that how to construct novel topological devices by utilizing the special surface states of 3D TIs.
Our method can provide a cheap and accurate numerical simulations
in these complicated structures, combined with the nonequilibrium Green's function formalism,
Landauer-B\"{u}ttiker formula, etc.
So the proposed method paves a new way to study the various properties of the surface states of the 3D TIs.

\section{\label{sec5}Conclusions}

In summary, we formulate a 2D lattice model for surface states of 3D TIs by appending a Wilson term with a zero bare mass to the massless Dirac equation.
The Wilson term can effectively ``solve'' the Fermi doubling problem by opening moderate gaps at the doubled Dirac cones and maintaining the one at $\Gamma$ point. The 2D lattice model provides a very coincident picture for both low-energy band structure and conductance of a TI nanowire with respect to a 3D model, but the numerical calculations based on 2D lattice model is much quicker than based on the 3D one. Moreover, the wormhole effect can be set up in a TI nanowire under a longitudinal magnetic field. We also find that the surface states of a TI nanowire are robust against disorder.

\section*{Acknowledgments}

{\bf Acknowledgments:}
This work was supported by NBRP of China (2015CB921102, 2014CB920901), NSF of China (11574007, 11274364, 11534001) and NSF
of Jiangsu Province, China (BK20160007).

\end{document}